\documentclass[manuscript]{copernicus}
\usepackage{tabularx}
\usepackage{multirow}
\usepackage{supertabular}
\usepackage{algorithmic}
\usepackage{algorithm}
\usepackage{amsthm}
\usepackage{float}
\usepackage{graphicx,psfrag,epsf}
\usepackage{enumerate}
\usepackage{natbib}
\usepackage{hyperref}
\usepackage{mathtools}

\begin{document}
\nolinenumbers

\title{A Robust Approach to Gaussian Processes Implementation}


\Author[1][julietnate@gmail.com]{Juliette}{Mukangango} 
\Author[2]{Amanda}{Muyskens}
\Author[2]{Benjamin}{W. Priest}

\affil[1]{Colorado School of Mines, 1500 Illinois St, Golden, CO 80401}
\affil[2]{Lawrence Livermore National Laboratory, 7000 East Ave, Livermore, CA 94550}




\runningtitle{A Robust Approach to Gaussian Processes Implementation}

\runningauthor{Juliette Mukangango}

\received{}
\pubdiscuss{} 
\revised{}
\accepted{}
\published{}


\firstpage{1}

\maketitle

\begin{abstract}
Gaussian Process (GP) regression is a flexible modeling technique used to predict outputs and to capture uncertainty in the predictions. However, the GP regression process becomes computationally intensive when the training spatial dataset has a large number of observations. To address this challenge, we introduce a scalable GP algorithm, termed \textit{MuyGPs}, which incorporates nearest neighbor and leave-one-out cross-validation during training. This approach enables the evaluation of large spatial datasets with state-of-the-art accuracy and speed in certain spatial problems. Despite these advantages, conventional quadratic loss functions used in the \textit{MuyGPs} optimization such as Root Mean Squared Error(RMSE), are highly influenced by outliers. We explore the behavior of \textit{MuyGPs} in cases involving outlying observations, and subsequently, develop a robust approach to handle and mitigate their impact. Specifically, we introduce a novel leave-one-out loss function based on the pseudo-Huber function (LOOPH) that effectively accounts for outliers in large spatial datasets within the \textit{MuyGPs} framework. Our simulation study shows that the "LOOPH" loss method maintains accuracy despite outlying observations, establishing \textit{MuyGPs} as a powerful tool for mitigating unusual observation impacts in the \textit{large data regime}. In the analysis of U.S. ozone data, \textit{MuyGPs} provides accurate predictions and uncertainty quantification, demonstrating its utility in managing data anomalies. Through these efforts, we advance the understanding of GP regression in spatial contexts.
\end{abstract}


\introduction  
Gaussian Process(GP) regression is widely known to be a powerful and versatile framework for modeling non-linear relationships in various fields. Particularly, in spatial data analysis, GP regression's capability to effectively account for the correlation among all data points makes it an attractive choice for interpolating highly non-linear targets \citep{Rasmussen2006, Cressie1993}. The versatile properties of GPs along with the native uncertainty quantification of their predictions has led to their adoption in different domains, such as time series forecasting \citep{Rasmussen2003}, and Bayesian optimization \citep{snoek2012}. Despite their advantages, the incorporation of GPs in the presence of outliers remains an open challenge. \cite{Stegle2008} address this challenge posed by noisy data, outliers, and missing information by proposing a model that combines unsupervised clustering and Bayesian regression. In the realm of robust GP regression, other researchers such as  \cite{jylanki2011, park2022, ranjan2016, li2021} have explored alternative approaches, including methods such as the Student-t likelihood, bias modeling, EM-based algorithms, and iterative trimming, and more.

Fundamentally in spatial data analysis, GP estimation aims to learn the covariance model hyperparameters (denoted as $\theta$) and use these estimates for interpolation with uncertainty quantification. However, the computational complexity of GP regression escalates with the presence of large sets of observations. GP regression scales cubically with the number of observations, rendering it impractical on large sets of observations. Further, it scales quadratically in memory, where traditional methods require formation and storage of the covariance matrix, which is prohibitive to their computation. Numerous studies, including those led by \cite{ForemanMackey2017, Csat2002, dietrich1997fast}, and others, have aimed to effectively address the challenge at hand. Methods such as Fixed Rank Kriging \citep{cressie2008}, Lattice Krig \citep{nychka2015}, Predictive Processes \citep{banerjee2008}, Vecchia \citep{katzfuss2021}, among others, have been proposed. For instance, the general Vecchia approach, as detailed by \cite{katzfuss2021}, represents a GP method that includes many popular approximations within a unified framework, leading to new insights regarding computational properties and allowing for significant improvements in approximation accuracy over Vecchia’s original method. Vecchia's method, while computationally feasible, may still involve intricate sparsity structures and complex implementations, potentially leading to less efficient hyperparameter optimization. The novel algorithm \textit{MuyGPs}, introduced by \cite{muygps2021}, stands out by achieving state-of-the-art fast prediction and hyperparameter optimization. \textit{MuyGPs}, by contrast, provides a simpler and more direct approach to maintaining prediction accuracy and computational efficiency. This superior performance of \textit{MuyGPs} has been demonstrated in various applications like those conducted by \cite{wood2022, Goumiri2022, Dunton2022, buchanan2022, muyskens2022, Goumiri2020}.


It is imperative to reinforce the \textit{MuyGPs} algorithm against potential outliers that could compromise the accuracy of the GP predictions and associated uncertainties. Outliers are present in numerous domains such as environmental data, where variables like air quality can exhibit outlying values. Researchers such as \cite{Knief2021} and \cite{Wang2019} demonstrate what happens when outliers are ignored and a GP is applied to data naively. The assumptions of the GP model are violated by the outliers, disrupting the underlying structure of the GP. Therefore the hyperparameter estimates and predictions are affected, leading to inaccurate inferences. These outliers offer larger errors, which makes them overly influential in loss functions based on squared error such as Root Mean Squared Error (RMSE). Additionally, these outliers introduce extra variability into the process and can lead to an overestimation of the variance, resulting in very wide prediction intervals. In our simulation study, fitting both conventional GP and Local Approximate GP (LAGP) regression models further illustrates and supports the points discussed in this paragraph. Addressing outliers systematically is essential not only for ensuring the robustness and accuracy of \textit{MuyGPs} but also for maintaining the integrity of GP methods in applications where outliers are prevalent. Exploring methodologies that explicitly account for outliers within the context of GP remains an ongoing area of interest for improving the reliability of predictions across diverse datasets and domains \citep{Stegle2008}.

To address concerns in both hyperparameter estimation and prediction from spatial datasets with outliers, we propose a refinement to the \textit{MuyGPs} optimization algorithm. First, we introduce a novel contribution called the leave-one-out pseudo-Huber (LOOPH) loss function. The LOOPH loss function combines ideas popularized by \cite{Huber1964} with a quadratic leave-one-out likelihood loss function inspired by the Gaussian likelihood defined by \cite{wood2022}. We show that LOOPH reduces the sensitivity of \textit{MuyGPs} optimization to large residuals resulting from outlying spatial data. Second, we explore augmenting the loss function by computing the central value such as the median metric of the distribution formed by repeatedly down-sampling nearest neighbor sets and evaluating the associated predictions. The median, as a robust measure of central tendency, provides a stable reference point less influenced by extreme values or outliers. Therefore, the median value is used in the optimization process to robustly tune hyperparameters, mitigating the algorithm's sensitivity to large residuals. \cite{reich2011} explored a modeling approach for skewed datasets by applying Quantile Regression (QR), but in cases where the focus is examining the mean and the overall trend in the presence of outliers, GP regression is a better option than QR.  Therefore, the refinement to the \textit{MuyGPs} algorithm address the challenges posed by noisy data, outliers, and missing information in GP models, ensuring accurate and reliable predictions across the full range of data points, not just the extremes, diverging from traditional extreme value methods discussed by \cite{Smith1989}.

\section{MuyGPs Uncertainty Calibration}
In this section, we outline in details the methodology behind the robust approach applied to the \textit{MuyGPs} algorithm. At the core of our methodology lies the integration of the variance-regularizing pseudo-Huber loss function, pioneered by \cite{filipovic2021}. This pivotal inclusion significantly enhances the algorithm's ability to manage outliers and noisy data, all while maintaining efficient scalability. We detail the integration process and highlight its significance in achieving accurate predictions across diverse datasets.

\subsection{Background: MuyGPs Overview}
Consider a spatial GP regression of the form:
\begin{equation}
    Y(\mathbf{x}) = \mathbf{x}^\top \beta + f(\mathbf{x}) + \varepsilon,
\end{equation}
where \( Y(\mathbf{x}) \) represents the spatial observation vector, \( \mathbf{x} \) is the feature matrix, \( \beta \) are the linear coefficients, \( f(\mathbf{x}) \) is  the underlying spatial function capturing the non-linear components, and \( \varepsilon \) accounts for the measurement noise. We define \( f(\mathbf{x}) \) as a GP if the function's values at any finite set of n points \(\mathbf{x} = (\mathbf{x_1},\ldots,\mathbf{x_n}) \) follow a multivariate normal distribution with mean zero and a covariance kernel \( K(\mathbf{x}_i, \mathbf{x}_j; \theta) \). That is, 
\begin{equation}
    f(\mathbf{x}) = (f(\mathbf{x_1}),\ldots,f(\mathbf{x_n}))^T \sim \mathcal{N}(\mathbf{0}, K(\mathbf{x}_i, \mathbf{x}_j; \theta)),
\end{equation}
where \( \mathbf{\theta}\) represents a set of hyperparameters controlling the behavior of the kernel. We parameterize a GP with a Mat\'ern Covariance function using a set of parameters $\mathbf{\theta} = (\sigma^2, \nu, \ell, \tau^2)^T$, where $\sigma^2$ is the scale parameter, $\nu$ is the smoothness parameter, $\ell$ represents the length-scale parameter, and $\tau^2$ denotes the homoscedastic measurement noise prior variance. We chose the Matérn covariance function over the Radial Basis function (RBF) and other covariance functions for its flexibility and ability to model environmental processes. According to \cite{Stein1999}, the Matérn covariance function is widely used in spatial modeling and represents the pointwise limit of smoothness in the RBF. The Matérn Covariance function is defined as:
\begin{equation}
    C_{\nu}(||\boldsymbol{h}||) = \sigma^2\frac{2^{1-\nu}}
{\Gamma(\nu)}\left( \sqrt{2\nu}\frac{||\boldsymbol{h}||}{\ell} \right)^{\nu}K_{\nu}\left(\sqrt{2\nu}\frac{||\boldsymbol{h}||}{\ell}\right),
\end{equation}
where $\mathbf{h}$ represents the distance between two locations, and $K_{\nu}$ is the modified Bessel function of the second kind. 
\\ Conventional GP training consists of maximizing the log-likelihood of the training data given $\theta$, which becomes very expensive in a large data regime. \cite{muygps2021} introduced \textit{MuyGPs} as a scalable GP regression algorithm designed specifically to address the challenges posed by large spatial datasets.

The methodology behind \textit{MuyGPs} is derived from the union of two concepts: 
\begin{itemize}
    \item \textit{Optimization with Leave-One-Out Cross-Validation}: By employing leave-one-out cross-validation, \textit{MuyGPs} avoids the need to evaluate expensive log-likelihood GP functions for each prediction. This optimization strategy significantly reduces computational costs, making \textit{MuyGPs} suitable for large spatial datasets where computational efficiency is crucial.
    \item \textit{Kernel Matrix Restriction to Nearest Neighborhood:} \textit{MuyGPs} restricts the kernel matrix to the k nearest neighbors of a prediction location. This restriction limits the cost of kriging weights, further enhancing computational efficiency without compromising model accuracy. Hence, \textit{MuyGPs} conditions $\mathbf{x}_i$ on its k nearest neighbors, denoted $\mathbf{X}_{N_i}$, yielding 
\begin{equation}
    \bar{\mu}_i = \hat{Y_{\theta}}(\mathbf{x}_i|\mathbf{X}_{N_i}) = K(\mathbf{x}_i, \mathbf{X}_{N_i};\theta) K(\mathbf{X}_{N_i}, \mathbf{X}_{N_i};\theta)^{-1} Y(\mathbf{X}_{N_i}), 
\end{equation}
\begin{equation}
    \bar{\Sigma}_{ii} = Var(\hat{Y_{\theta}}(\mathbf{x}_i|\mathbf{X}_{N_i}) )=  K(\mathbf{x}_i, \mathbf{x}_i ;\theta) - 
 K(\mathbf{x}_i, \mathbf{X}_{N_i};\theta) K(\mathbf{X}_{N_i}, \mathbf{X}_{N_i};\theta)^{-1} K( \mathbf{X}_{N_i}, \mathbf{x}_i;\theta)
\end{equation} 
as the predictors of the response distribution. 
\end{itemize}
While other researchers may have explored the above concepts in different ways, \textit{MuyGPs} method is the first to leverage both insights simultaneously to accelerate kernel hyperparameter estimation by enforcing sparsity in the kriging weights. This sparsity not only speeds up training but also improves the scalability of \textit{MuyGPs} for handling massive spatial datasets. 

The \textit{MuyGPs} training process then minimize several loss functions such as the Mean Squared Error (MSE), the cross entropy loss, and the leave-one-out likelihood (LOOL) loss over randomly sampled batch of training points \textbf{B}. These loss functions play a significant role in defining the objective function for hyperparameter optimization during training, thereby having a big impact on the training results. Because \textit{MuyGPs} method's success depends on the combination of the leave-one-out cross-validation and nearest-neighbor approximations, we naturally choose the LOOL loss function as a primary loss function for the optimization process. LOOL loss function lets us use both the above features while keeping our predictions accurate and varied. 

For a randomly-selected training batch $\mathbf{B}$ with b elements, the hyperparameters $\mathbf{\theta}$ minimize the following loss function:
\begin{equation}
    Q(\mathbf{\theta}) = \sum_{i \in \mathbf{B}}  \left(\frac{(\bar{\mu}_i - y_i)^2}{\bar{\Sigma}_{ii}} + \log \bar{\Sigma}_{ii}\right),
\end{equation}
where $\bar{\mu}_i$ and $\bar{\Sigma}_{ii}$ are the posterior mean and variance of the $i^{th}$ batch point, as defined in Eq. (4) and Eq. (5), respectively. 
The loss function, denoted as $Q(\mathbf{\theta})$, is referred to as the LOOL loss function \citep{wood2022}, which in typical scenarios achieves favorable performance for \textit{MuyGPs}. However, its efficacy can be compromised when the spatial dataset at hand includes outliers. In the subsequent subsection, we provide a comprehensive breakdown of the integration process of the variance regularized robust function into the \textit{MuyGPs} algorithm.

\subsection{Robust Process}
In the context of GP regression, addressing outliers is pivotal for ensuring model robustness and reliable predictions. To tackle this challenge, we turn to the pseudo-Huber loss function, which has garnered recognition for its effectiveness in reducing the impact of outliers, \citep{filipovic2021}. This loss function serves as a smooth approximation to the Huber loss (\cite{huber1992}), a widely-known method for handling outliers in various statistical and machine learning applications. The pseudo-Huber loss function is  defined as follows:
\begin{equation}
 \sum_{i=1}^b \delta^2\left(\sqrt{1+\left(\frac{\bar{\mu}_i - y_i}{\delta}\right)^2} - 1\right),
\end{equation}
where $\delta$ is the boundary scale parameter which controls the amount of robustness of the loss function and $\bar{\mu}_i$ and $y_i$ are the same quantities in Eq. (6). Figure 1 illustrates that the pseudo-Huber loss displays quadratic behavior for small residuals and linear behavior for larger residuals, depending on the chosen $\delta$ value. This dual nature makes it less sensitive to the influence of outliers.

\begin{figure}[H]
    \centering  \includegraphics{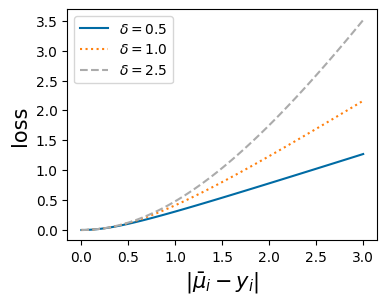}
    \caption{The pseudo-Huber loss function for different boundary scale values. Each curve represents a different boundary scale, demonstrating the transition from quadratic behavior for small residuals to linear behavior for large residuals.}
    \label{fig:f01}
\end{figure}

To enhance the loss function's sensitivity to variance-affecting parameters, we introduce a novel method, the LOOPH. This method scales and regularizes the pseudo-Huber loss, ensuring that it reacts more strongly to parameters influencing variance. The formulation of the LOOPH is as follows:
\begin{equation}
\sum_{i=1}^b 2\delta^2\left(\sqrt{1+\frac{(\bar{\mu}_i - y_i)^2}{\delta^2 \bar{\Sigma}_{ii}}} - 1\right)+ \log \bar{\Sigma}_{ii}.
\end{equation}
The LOOPH loss explicitly depends on the posterior variance, $\bar{\Sigma}_{ii}$; which makes the loss more sensitive to variance parameters due to penalizing large variances. The constant $\delta$ is auto-normalized to set the boundary between quadratic and linear behavior of the residual. This addresses one of the major problems with the Huber and pseudo-Huber losses by assigning directly interpretable units to the boundary scale parameter. In the LOOPH loss function, the boundary scale parameter is interpreted as the number of standard deviations from the posterior mean beyond which residual losses are linearized. In our analysis, we selected $\delta=3$ to ensure that penalties for residuals beyond three standard deviations from the target are linearized. This choice aligns with the common assumption that approximately $99.7\%$ of data falls within this range under the normality assumption. Setting $\delta=3$ strikes a suitable balance for most applications and provides a standardized approach without requiring frequent adjustments. In their works, \cite{ronchetti2009} and \cite{rousseeuw2005} discuss the rarity of data points greater than three standard deviations away in a normal distribution, supporting the use of this cutoff as an effective outlier criterion. 

\begin{figure}[H]
    \centering    
    \includegraphics[width = \textwidth]{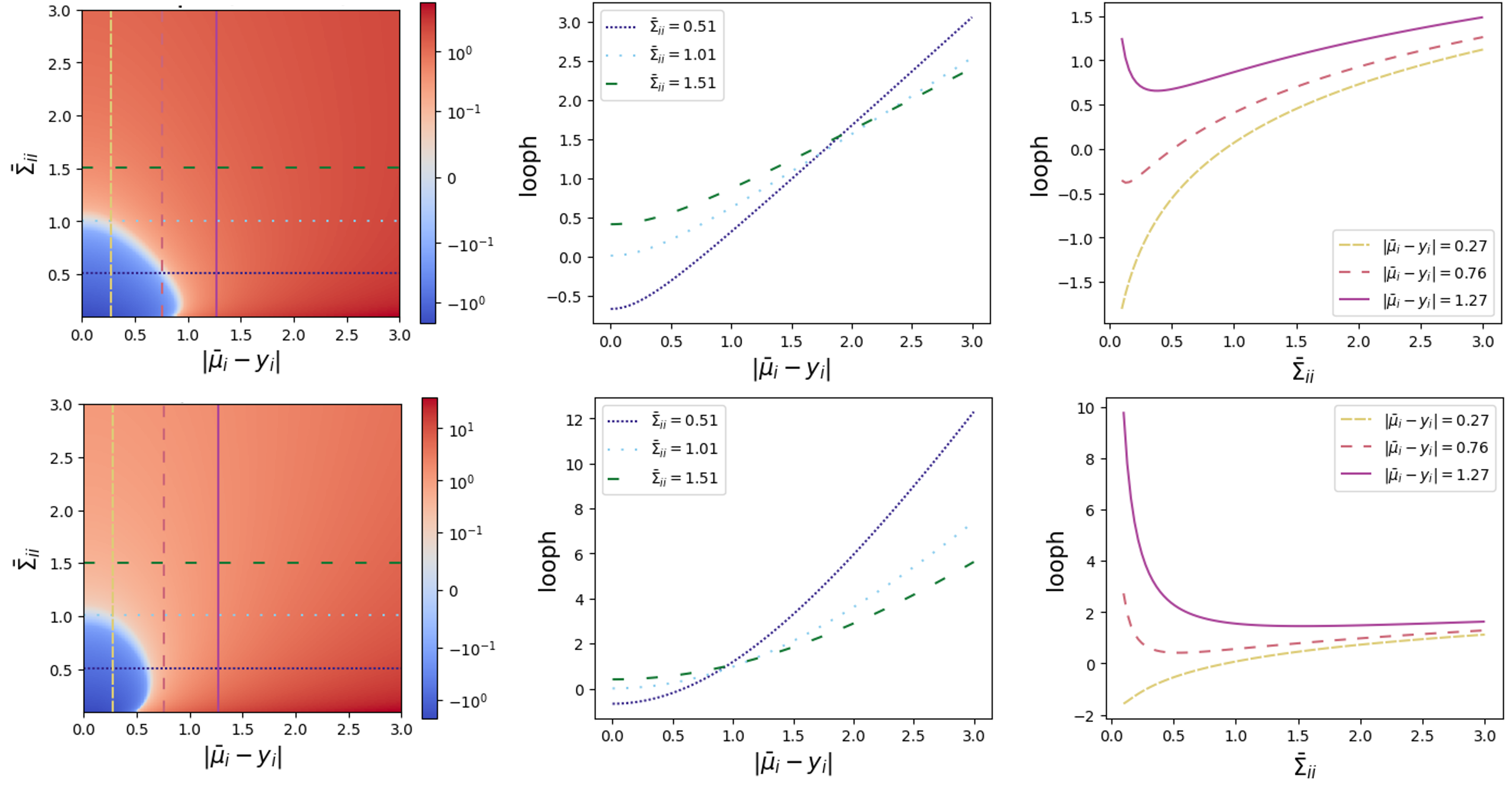}
    \caption{ Plots of the LOOPH function (left column), the LOOPH function against residuals (middle column), and the LOOPH function against variance (right column) for $\delta = 0.5$ (upper row) and $\delta = 3.0$ (lower row).
    }
    \label{fig:f02}
\end{figure}

Our comprehensive set of visualizations in Fig. 2 includes heatmaps that depict the loss surface across a range of $\delta$, $\bar{\mu}_i - y_i$, and $\bar{\Sigma}_{ii}$ values. These heatmaps provide a two-dimensional (2-D) view of the loss surface, with colors representing loss magnitude. Additionally, we employ cross-sectional line plots to showcase how the loss changes with varying residuals for different values of $\bar{\Sigma}$ and vice versa. The graphs presented in Fig. 2 illustrate that the steepness of the loss curve increases notably when the residuals get larger. However, this heightened slope doesn't surpass the impact of variability within the loss. For values of $\delta$ that are too small, as exemplified by the top row of Fig. 2, the loss attributed to the residual becomes excessively linearized. This tendency encourages an overestimation of the variance. Importantly, in practical scenarios, both aspects of the LOOPH loss may necessitate more training iterations to converge to a stable solution compared to alternative methods.
\begin{figure}[H]
    \centering    \includegraphics[width=\textwidth]{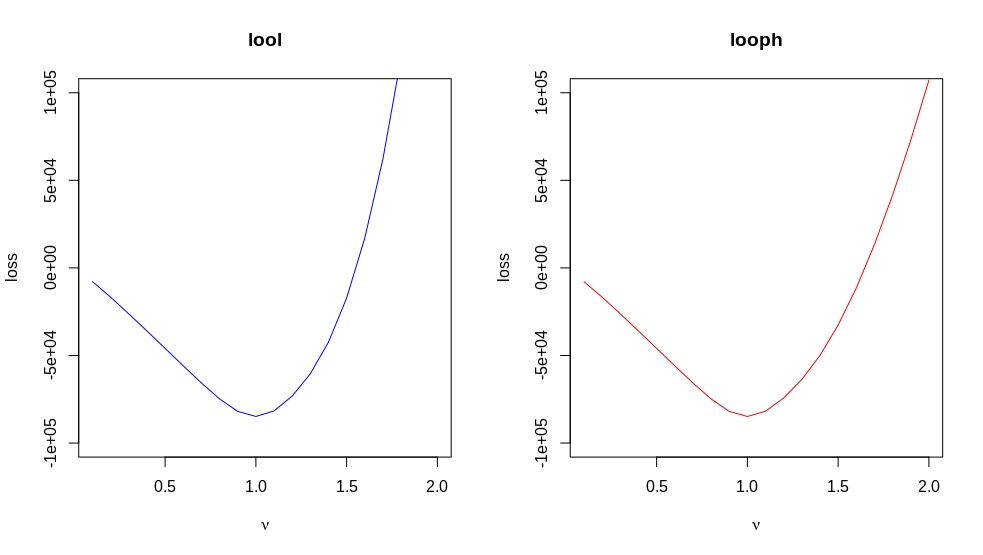}
    \caption{Comparison of loss functions for different values of $\nu$. The left plot shows the LOOL loss against various $\nu$ values, while the right plot displays the LOOPH loss against the same $\nu$ values.}
    \label{fig:f03}
\end{figure}
To compare ``LOOL'' and ``LOOPH'' loss functions, we tested them with different $\nu$ values while keeping $\delta$ fixed at three. When we did this comparison, we found that even when there were no outliers in the dataset, the ``LOOPH'' optimization surface remained steeper than the ``LOOL'' surface, as showcased in Fig. 3. The difference in steepness is significant because it reflects the sensitivity of the optimization process to changes in the input, highlighting that ``LOOPH'' responds more swiftly to variations in $\nu$ values. This heightened responsiveness can impact the model's adaptability and overall performance. Additionally, we found faster convergence empirically for simple problems with no outliers.
\subsection{Down-Sampling Process}
To enhance the robustness and predictive reliability of the \textit{MuyGPs} algorithm, we consider an innovative strategy that incorporates down-sampling of nearest neighbors and repeated evaluations to derive a central value of the distribution such as the median. This technique aims to strengthen the stability and accuracy of the \textit{MuyGPs} training and prediction process, particularly when confronted with outliers and other perturbations.  In our analysis, we assume that we know the true $\ell$ parameter value and we estimate $\nu$ parameter. $\sigma^2$ as an additional hyperparameter is treated differently than others and we separately optimize it by invoking a function based upon the mean of the closed-form $\sigma^2$ solutions associated with each of its batched nearest neighbor sets. We now illustrate a few steps at the heart of the down-sampling method to train $\nu$:
\begin{algorithm}[H]
    \caption{Down-Sampling Algorithm}
    \begin{enumerate}
        \item Begin by selecting the $k$-nearest neighbor counts, batch count, and down-sampling size, and then sample batches of data.
        \item Down-sample the nearest neighbor points and compute the objective function using Bayesian optimization to obtain predictions for the $\nu$ parameter. Repeat this process for a fixed number of iterations and obtain the median value of the associated distribution.
        \item Fix the robust $\nu$ obtained from the median metric in the optimization and prediction process for accurate and robust inference.
        \item Estimate the $\sigma^2$ parameter using the down-sampled indices of the nearest neighbors to minimize variability in inference. Compute $\hat{\sigma}^2$ as follows:
        \begin{equation}
            \hat{\sigma}^2 = \frac{1}{b^*k^*}\text{med}(Y^T_{{nn}^*}K^{-1}_{{nn}^*}Y_{{nn}^*}),
        \end{equation}
        where $Y_{{nn}^*}$ and $K_{{nn}^*}$ are the target and kernel matrices with respect to the down-sampled nearest neighbors. $k^*$ is the subset of the nearest neighbor counts, and $b^*$ is the subset of the batch count.
        \item Down-sample the nearest neighbors again to predict the response of the test data at a fixed number of iterations. Each iteration results in a distribution of predictions. Compute the median of these predictions to obtain a central value, contributing to robustness and predictive reliability, especially when dealing with outliers and other sources of variability.
    \end{enumerate}
\end{algorithm}

The validation of the described robust approach will be presented in the subsequent section, where we delve into the numerical results and performance analysis. We will then test a \textbf{Hybrid} method where we only use the down-sample strategy for $\sigma^2$ parameter and use the full batch to estimate all other parameters.


\section{Numerical Studies}
To assess the effectiveness of the proposed LOOPH method and  batch sub-sampling technique, we conducted a series of experiments using a simulated dataset as well as a real dataset. Throughout these experiments, we followed the structure below:
\begin{itemize}
    \item We fitted \textit{MuyGPs} models using three methods: 
    \begin{itemize}
        \item \textit{Regular Sampling method:} which is the traditional \textit{MuyGPs} implementation.
        \item \textit{Hybrid method:} which involves the down-sampling of nearest neighbor indices only for $\sigma^2$. 
        \item \textit{Down-Sampling method:} see Alg 1.
    \end{itemize}
    \item For each \textit{MuyGPs} model we applied the LOOL (Eq. 6) and the LOOPH (Eq. 8) loss functions.
    \item We fitted three additional models that employ the negative log-likelihood (NLL) as the underlying loss function. The models are as follows:
    \begin{itemize}
        \item \textit{Conventional GP:} a conventional GP model fit using ``Fields'' R package, \citep{nychka2021}.
        \item \textit{LAGP}: a local approximate GP regression model for large spatial datasets.
        \item \textit{Student-t GP}: a GP regression model with a Student-t likelihood, see \cite{jylanki2011}. 
    \end{itemize}
    \item For all of the models, we used pure (non-outlying) and outlying data for comparison and varied three $\nu$ values.
\end{itemize}

\subsection{Simulation Experiment}
The simulation study employs a simple 2-D curve generated from $f(\mathbf{x})$. The domain is defined as a simple grid on a one-dimensional surface, and the observations are partitioned into a $90\%$ training set and a $10\%$ test set. Each dimension of the dataset consists of 100 data points, which results in a data size of 10,000 observations. We assume that the true data is produced with no noise, so we specify a very small noise prior to ensure numerical stability, i.e a minimal noise level of $1 \times 10^{-14}$. The training observations are perturbed with heteroscedastic Gaussian noise with variance $\varepsilon = 1 \times 10^{-7}$. The kernel hyperparameters, including different values of the smoothness parameter, i.e $\nu = 0.1,0.5,1.0$ and the length scale parameter $\ell = 1.0$, are then specified. These parameters collectively define a Mat\'ern kernel GP for our sampling procedure. Figure 4 displays our sampled surface from the GP prior and shows the training and testing observations.

\begin{figure}[H]
    \centering   
    \includegraphics[width = \textwidth, height = 5cm]{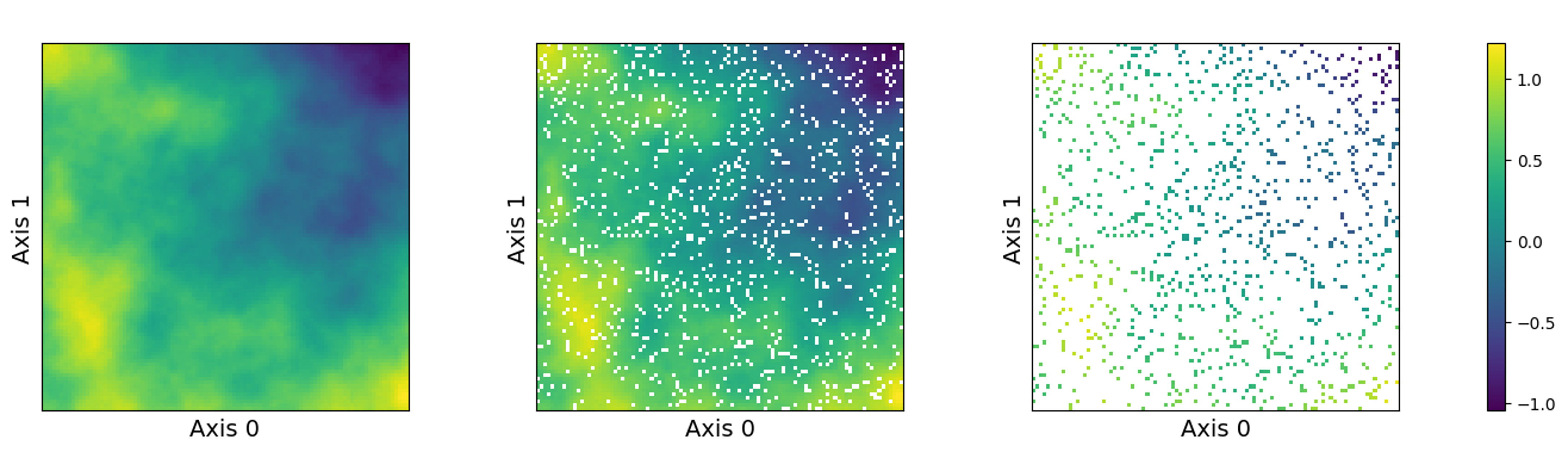}
    \caption{From left to right panel: 2-D sampled surface from a GP, $90\%$ training observations, and $10\%$ testing observations.}
    \label{fig:f04}
\end{figure}

In order to investigate the robustness of our model to outliers, we introduce anomalous data points into the training set. This is achieved by randomly selecting a subset of training data indices and multiplying the corresponding target values by a specified factor. Specifically, we randomly choose $10\%$ of the training data points and scale their target values by a factor of two. This operation effectively injects outliers into the dataset, simulating situations where extreme observations may exist as seen in Fig. 5 and Fig. 6. Particularly, in Fig. 5 the two box plots illustrate distinct characteristics of the training datasets they represent. The first box plot displays a tightly clustered distribution of values with no outliers, indicating a more consistent and predictable dataset. In contrast, the second box plot reveals a wider spread of values and the presence of outliers, highlighting the dataset's increased variability and potential for extreme observations. These plots underscore the significance of outliers in data analysis, emphasizing the need for careful consideration when interpreting results or applying statistical methods to datasets with outliers.
\begin{figure}[H]
    \centering  
    \includegraphics[width = 7cm, height = 3cm]{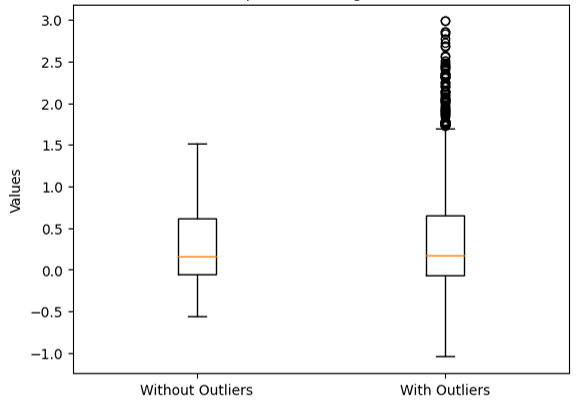}
    \caption{Comparison of the spatial training data after injecting outliers.  The first boxplot represents the data with no outliers, while the second boxplot shows the data after injecting $10\%$ outliers.}
    \label{fig:f05}
\end{figure}

\begin{figure}[H]
    \centering    
    \includegraphics[width = 11cm, height = 9cm]{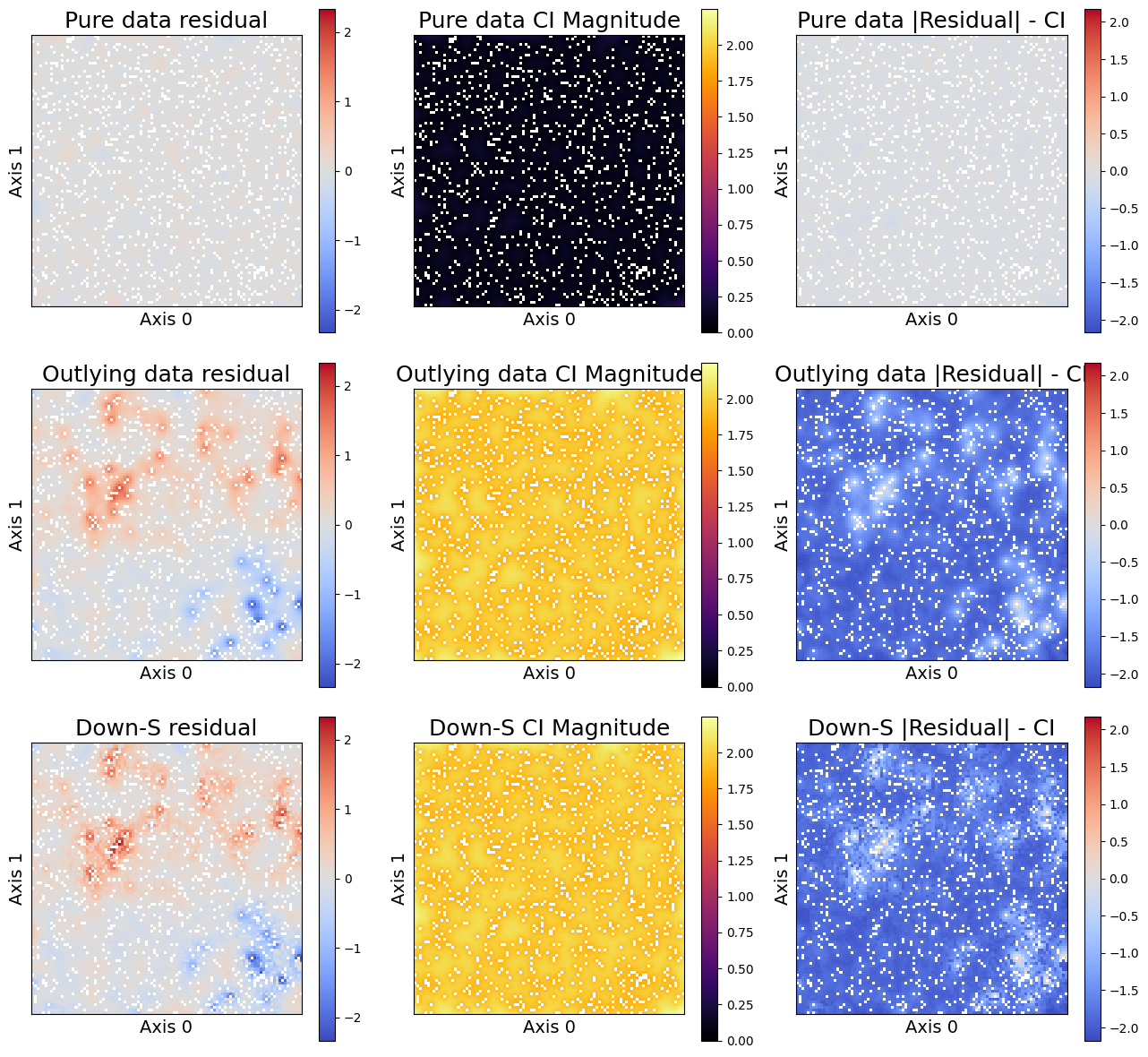}
    \caption{Analysis of outlier effects across three models. The left column depicts residuals, the middle column shows the confidence interval (CI) magnitude, and the right column illustrates the difference between the residuals and CI (|Residual| - CI). The first row presents a model trained on pure data with no outliers using the regular sampling method with LOOL loss. The second row shows a model trained on data with $10\%$ outliers using the regular sampling method with LOOL loss. The third row displays a model trained on data with $10\%$ outliers using the Down-Sampling approach with LOOL loss. The comparison highlights the impact of outliers on different MuyGPs models.}
    \label{fig:f06}
\end{figure}

The plots illustrated in Fig. 6 serve as valuable tools for gaining insights into our findings regarding outlier effects. In the left column, we analyze the residuals computed from three models. Notably, the model trained on data with outliers exhibit considerably larger residuals, which could potentially impact the validity of our inferences. In the middle column, we examine the size of the $95\%$ confidence interval calculated for all the models. The presence of outliers tends to significantly widen the $95\%$ confidence interval, indicating decreased confidence in the model's predictions. The average confidence interval size in the model with outlier-affected data is approximately 2.25, notably higher compared to the average size observed in the model with outlier-free data, averaging around 0.55. Lastly, the right column illustrates the difference between the $95\%$ confidence interval length and the magnitude of residuals for all the models. Any points exceeding zero lie outside the confidence interval, providing insight into our coverage distribution. While approximately $95\%$ of the differences in confidence intervals and residuals for the outlying data do not exceed zero as expected, the majority are significantly negative. This suggests that the learned confidence intervals are made excessively large to accommodate the outlying data. Overall, these results highlight the significant impact outliers can have on model performance and the importance of robust methods to mitigate these effects. For clear visual interpretation, we specifically report plots of predictions made using larger test sets.
\subsection{Simulation Results}
The simulation results are all based on $100$ replications, and for each simulated data set we computed the RMSE, the continuous ranked probability score (CRPS), the median absolute deviation (MAD), the median diagonal variance (MDV), the median confidence interval (CI) size, and the coverage probability to analyze our fitted models. Table 1  summarize the key findings from our simulation study in the absence of outliers. For brevity, we report only RMSE, MDV, and median CI size; the rest are presented in the appendix. RMSE was chosen because it provides a clear measure of model accuracy, showing how closely the predictions match the true values. MDV is useful for assessing the variability in predictions, offering insights into the stability of the model. Median CI size is essential for understanding the uncertainty in the predictions, indicating the precision of the model's estimates. While MDV measures how spread out the predictions are, reflecting the model's consistency, median CI size evaluates the range within which the true values are likely to fall, highlighting the confidence in the predictions. Thus, MDV and CI size provide complementary information about the model's performance. The bold values represent the best accuracy and uncertainty quantification (UQ) statistics for each $\nu$ setting.

\begin{table}[H]
    \centering
    \begin{tabular}
    {|c|c|c|c|c|c|c|}
    \hline
       Model & Method  & Loss & $\nu$ & RMSE & MDV & CI Size \\
       \hline
        \multirow{18}{7em}{MuyGPs} & Regular Sampling & LOOL & $0.1$ & $0.59$ &$0.349$ & $2.31$ \\
        &&& $0.5$ & $0.089$  & $0.0077$ & $0.344$ \\
        &&& $1.0$ & $0.009$ & $8.0e^{-5}$ & $0.034$\\
        \cline{3-7}
        && LOOPH & $0.1$ & $0.581$ & $0.342$ & $2.293$ \\
        &&& $0.5$ & $0.0887$ & $0.007$ & $0.336$  \\
        &&& $1.0$ & $0.0092$ & $\mathbf{6.8e^{-5}}$ & $\mathbf{0.032}$ \\
        \cline{2-7}
         & Hybrid & LOOL & $0.1$ & $0.612$ & $0.344$ & $2.29$\\
        &&& $0.5$ & $0.095$ & $0.0078$ & $0.347$ \\
        &&& $1.0$ & $0.0098$ & $0.0001$ & $0.035$ \\
        \cline{3-7}
        && LOOPH & $0.1$ & $0.6002$ & $\mathbf{0.319}$ & $\mathbf{2.216}$ \\
        &&& $0.5$ & $0.089$ & $\mathbf{0.0068}$ & $\mathbf{0.324}$ \\
        &&& $1.0$ & $0.0096$ & $7e^{-5}$ & $0.0326$ \\
        \cline{2-7}
        & Down-Sampling & LOOL & $0.1$ & $0.611$ & $0.397$ & $2.47$ \\
        &&& $0.5$ & $0.089$ & $0.0082$ & $0.355$ \\
        &&& $1.0$ & $0.0096$ & $0.0001$ & $0.038$ \\
        \cline{3-7}
        && LOOPH & $0.1$ & $0.603$ &  $0.329$ & $2.25$\\
        &&& $0.5$ & $0.089$ & $0.0075$ & $0.34$\\
        &&& $1.0$ & $0.0095$ & $8e^{-5}$ & $0.035$  \\
        \hline
        \multirow{9}{7em}{Benchmarks} & Conventional GP & NLL & $0.1$ & $\mathbf{0.017}$ & $0.475$ & $2.703$\\
        &&& $0.5$ & $\mathbf{0.003}$ & $0.883$ & $3.683$\\\
        &&& $1.0$ & $\mathbf{0.0005}$ & $0.523$ & $2.836$\\
        \cline{2-7}
        & LAGP & NLL & $0.1$ & $0.947$ & $0.704$ & $3.2895$\\
        &&& $0.5$ & $0.958$ & $0.214$ & $1.812$\\
        &&& $1.0$ & $0.725$ & $0.032$ & $0.703$\\
        \cline{2-7}
        & Student-t GP & NLL & $0.1$ & $0.944$ & $0.456$ & $5.82$ \\
        &&& $0.5$ & $0.747$ & $0.385$ & $5.099$\\
        &&& $1.0$ & $0.349$ & $0.225$ & $3.875$\\
        \hline
    \end{tabular}
   
   \caption{Results for model evaluation using RMSE, MDV, and median CI size metrics for non-outlying data. The $\nu$ column represents the true $\nu$ values under which the simulated data were generated.}
    \label{tab:t1}
\end{table}

We next summarize the results of our simulation study in Table 2, demonstrating our models' effectiveness in capturing underlying data patterns and their robustness in handling outliers. 
\begin{table}[H]
    \centering
    \begin{tabular}{|c|c|c|c|c|c|c|}
    \hline
       Model & Method  & Loss & $\nu$ & RMSE & MDV &  CI Size\\
       \hline
        \multirow{18}{7em}{MuyGPs} & Regular Sampling & LOOL & $0.1$ & $0.625$ & $0.569$ & $2.957$\\
        &&& $0.5$ & $0.206$ & $0.243$ & $1.93$\\
         &&& $1.0$ & $0.21$ & $0.278$ & $2.065$ \\
         \cline{3-7}
         && LOOPH & $0.1$ & $0.618$ & $0.475$ & $2.704$\\
         &&& $0.5$ & $\mathbf{0.126}$ & $\mathbf{0.044}$ & $\mathbf{0.819}$\\
         &&& $1.0$ & $\mathbf{0.058}$ & $0.031$ & $0.685$ \\
         \cline{2-7}
         & Hybrid  & LOOL & $0.1$ & $0.6012$ & $0.527$ & $2.846$ \\
         &&& $0.5$ & $0.176$ & $0.119$ & $1.35$ \\
         &&& $1.0$ & $0.227$ & $0.292$ & $2.118$ \\
         \cline{3-7}
         && LOOPH & $0.1$ & $0.606$ & $\mathbf{0.468}$ & $\mathbf{2.682}$ \\
         &&& $0.5$ & $0.155$ & $0.061$ & $0.971$ \\
         &&& $1.0$ & $0.085$ & $0.023$ & $0.593$ \\
         \cline{2-7}
         & Down-Sampling & LOOL & $0.1$ & $0.619$ & $0.488$ & $2.74$ \\
         &&& $0.5$ & $0.143$ & $0.067$ & $1.016$\\
         &&& $1.0$ & $0.0587$ & $\mathbf{0.014}$ & $\mathbf{0.466}$\\
         \cline{3-7}
         && LOOPH & $0.1$ & $\mathbf{0.6011}$ & $0.506$ & $2.78$ \\
         &&& $0.5$ & $0.15$ &$0.055$ & $0.922$ \\
         &&& $1.0$ & $0.062$ & $0.016$ & $0.503$\\
         \hline
         \multirow{9}{7em}{Benchmarks} & Conventional GP & NLL & $0.1$ & $0.941$ & $0.869$ & $3.656$\\
         &&& $0.5$ & $0.998$ & $1.609$ & $4.972$\\
         &&& $1.0$ & $0.869$ & $0.968$ & $3.857$\\
         \cline{2-7}
         & LAGP & NLL & $0.1$ & $1.027$ & $2.697$ & $6.438$\\
         &&& $0.5$ & $1.14$ & $3.958$ & $7.799$\\
         &&& $1.0$ & $0.998$ & $3.217$ & $7.03$\\
         \cline{2-7}
         & Student-t GP & NLL & $0.1$ & $0.9411$ & $0.846$ & $7.92$\\
         &&& $0.5$ & $0.927$ & $0.885$ & $8.85$\\
         &&& $1.0$ & $0.869$ & $0.769$ & $7.465$\\
         \hline
    \end{tabular}
    \caption{Results for model evaluation using RMSE, MDV, and median CI size metrics for data with $10\%$ outliers in it. The $\nu$ column represents the true $\nu$ values under which the simulated data were generated. }
    \label{tab:t2}
\end{table}

Examining the outcomes presented in Table 1 $\&$ Table 2 reveals several noteworthy insights. Remember that traditional implementation of the \textit{MuyGPs} method is noted here as ``Regular sampling'' with the ``LOOL'' loss function. All other rows indicate a novel method of our proposing with either a new loss function or novel use of data in order to account for the outliers,  along with conventional GP methods and one existing robust method for comparison. Our models exhibit exceptional accuracy when applied to clean data for both loss methods. This is evident in the form of low RMSE especially when $\nu = 1.0$, indicating precise point predictions. We can observe small MDV and precise confidence intervals when looking at the results in Table 1 for all the \textit{MuyGPs} methods. Surprisingly, our ``LOOPH'' loss model even outperforms the ``LOOL'' loss function in this clean data case in a majority of these statistics and scenarios. In contrast, the conventional GP method, evaluated with NLL loss, achieves the lowest RMSE values across all $\nu$ levels but at the cost of higher MDV and CI sizes. LAGP and Student-t GP methods show higher RMSE values, particularly at lower $\nu$ levels, with LAGP achieving lower CI sizes than Student-t GP. Overall, the Hybrid method with looph loss stands out due to its performance with low MDV and CI sizes.
\\The introduction of outliers has a profound impact on \textit{MuyGPs} models' accuracy when using the "LOOL" loss method, resulting in significantly enlarged variances and broader, overly conservative confidence intervals. RMSE values are higher than in Table 1 due to the influence of outliers, but they remain commendably low, particularly when $\nu = 1.0$ for all of our methods. Conversely, when assessing the results obtained using the "LOOPH" loss method in the presence of outliers, we still observe a small MDV and small median CI sizes. The "Down-Sampling" method exhibits improved robustness to outliers for the "LOOL" loss, delivering competitive inferential outcomes regardless of the presence of outliers. This is highlighted by substantially lower RMSE, diminished MDV values, and accurate confidence intervals. In contrast, the conventional GP and LAGP methods present significantly wider confidence intervals and increased MDV values when outliers are present, likely due to their sensitivity to outliers and the resultant increase in uncertainty. This increased variability and broader CI sizes for these methods indicate less stable performance, with RMSE values also notably higher, showcasing their reduced robustness compared to methods specifically designed to handle outliers. Although the Student-t GP method is robust to outliers, it still couldn't measure up to our developed methods in terms of maintaining low RMSE and variability. Our simulation analysis, supported by the conclusions derived from the earlier-mentioned tables, is further reinforced by the information presented in Fig. 6. 

\subsection{Analysis of U.S Ozone data}
In this subsection, we analyze the U.S. air quality data from various locations within Los Angeles (LA), CA in 1988. We considered the region's historical ozone levels, which have been notably high due to its status as a large metropolitan area. Throughout the 1980s and 1990s, LA recorded ozone levels exceeding 200 parts per billion (ppb). Although this dataset does not contain significant outliers, it is still critical to use a robust approach for accurate environmental analysis to account for potential future outliers that could be caused by climate change. Ozone levels are typically influenced by numerous factors, including weather patterns, emissions from various sources, and chemical reactions in the atmosphere. Our robust modeling approach ensures that the analysis remains reliable even when data variability is high or when there are subtle anomalies that traditional methods might overlook. By applying our robust techniques, we can better account for the complex nature of this dataset and improve the reliability of predictions and interventions aimed at mitigating air pollution.

We collected meteorological data from the National Climate Data Center (NCDC), providing $1096$ daily records of average temperature and maximum wind speed from three monitoring stations in LA. Additionally, we obtained maximum daily eight-hour average ozone levels from the US Environmental Protection Agency (EPA)'s Air Explorer Database. Our analysis primarily focused on daily ozone concentrations recorded at 15 monitoring sites in LA, totaling $6995$ observations, as shown in Fig. 7. To narrow down the datasets, we filtered for the summer months (June, July, and August), resulting in $1799$ ozone and $276$ meteorological observations.
\begin{figure}[H]
    \centering    
    \includegraphics[width = \textwidth]{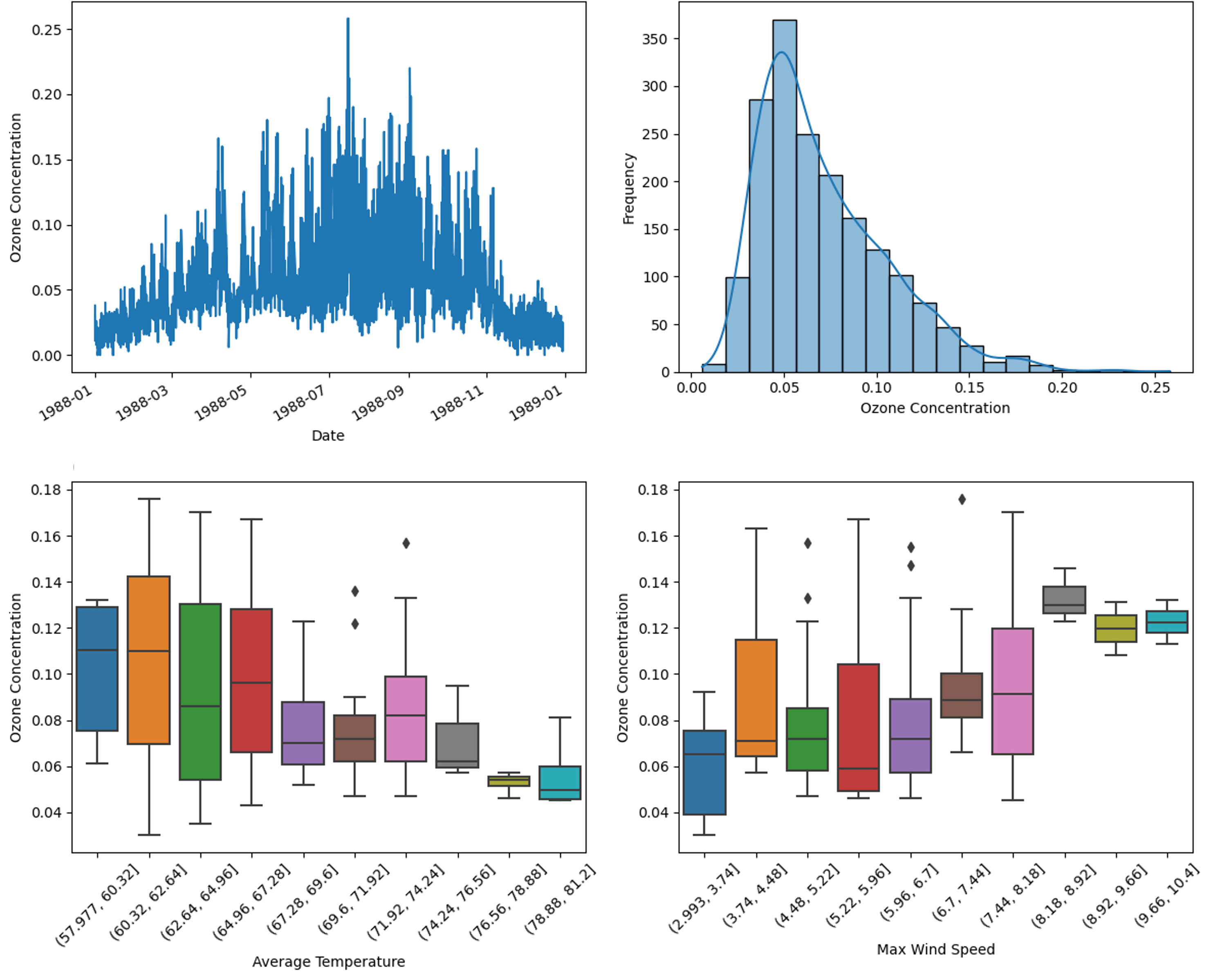}
    \caption{The upper left panel displays the time series of ozone concentration over time. The upper right panel shows the histogram of summertime ozone concentrations. The lower panel plots ozone concentration against daily average temperature and maximum wind speed.}
    \label{fig:f07}
\end{figure}

We followed a modeling approach almost similar to our simulation study, using temperature and wind speed as variables in our feature matrix and daily ozone concentration as our target variable. The feature variables were normalized using the min-max scaling to transform the values to a range of $[0,1]$. However; in the Ozone data analysis, we only estimated the $\nu$ parameter during the model training process due to the complex underlying characteristics and structures of this data and opted out of specifying its values. We considered Ozone values above the air quality standard as potential outliers. However, these values were not extreme enough to significantly affect the accuracy of our inference when using the "LOOL" loss method. Therefore, we randomly selected $10\%$ of the training Ozone data and replaced their values with extreme outlier values within a specified range to thoroughly test our methodology. Based on our simulation study, which showed that conventional GP, LAGP, and Student-t GP methods do not perform well due to scalability issues and lack of robustness (with only the Student-t GP being somewhat robust), we opted out of fitting these methods in the analysis of U.S Ozone data. Instead, we compared our method's performance to the regular \textit{MuyGPs} method by computing the RMSE, CRPS, MAD, MDV, median CI size, and coverage probability. Tables 3 and 4 only present RMSE, MDV, and median CI size metrics, while the rest of the metrics are included in the appendix. The results presented in these tables confirmed our earlier observations from the simulation study, favoring the "LOOPH" loss as well as the down-sampling approach.

\begin{table}[H]
    \centering
    \begin{tabular}{|c|c|c|c|c|}
    \hline
       Method  & Loss & RMSE & MDV & CI Size \\
       \hline
        \multirow{2}{5em}{Regular Sampling} & LOOL & $0.034$ & $8.6e^{-4}$ & $0.115$\\
         & LOOPH & $0.0336$ & $\mathbf{7e^{-5}}$ & $\mathbf{0.033}$ \\
         \hline
         \multirow{2}{5em}{Hybrid} & LOOL & $0.0336$ &  $3.51e{-3}$ & $0.232$ \\
         & LOOPH & $0.0336$ & $0.001$ & $0.127$ \\
         \hline
         \multirow{2}{5em}{Down-Sampling} & LOOL & $\mathbf{0.033}$ & $0.0005$ & $0.086$\\
         & LOOPH & $0.0339$ & $9e^{-5}$ & $0.0379$ \\
         \hline
    \end{tabular}
    \caption{Results for \textit{MuyGPs} models evaluation using RMSE, MDV, and Median CI size.}
    \label{tab:t3}
\end{table}
\begin{table}[H]
    \centering
    \begin{tabular} {|c|c|c|c|c|}
    \hline
       Method  & Loss & RMSE & MDV & CI Size \\
       \hline
        \multirow{2}{5em}{Regular Sampling} & LOOL & $0.5412$ &  $1.599$ & $4.95$ \\
         &  LOOPH & $0.489$ & $0.114$ & $1.323$\\
         \hline
        \multirow{2}{5em}{Hybrid} & LOOL & $0.54$ & $2.611$ & $6.334$ \\
         & LOOPH & $\mathbf{0.489}$ & $\mathbf{0.094}$ & $\mathbf{1.204}$ \\
         \hline
         \multirow{2}{5em}{Down-Sampling} & LOOL & $0.556$ & $0.149$ & $1.51$ \\
         & LOOPH & $0.509$ & $0.156$ & $1.548$ \\
         \hline
    \end{tabular}
    \caption{Results for \textit{MuyGPs} models evaluation using RMSE, MDV, and Median CI size for Ozone data with $10\%$ outliers generated in it.}
    \label{tab:t4}
\end{table}

The above findings in Table 3 $\&$ Table 4 provide a comprehensive view of the performance evaluation metrics for different batch sampling methods. It demonstrates how the models respond to various conditions, including the presence of outliers and the choice of loss functions (``LOOL'' and ``LOOPH''). Generally, these results highlight the trade-offs between accuracy and robustness in GP modeling. Even without injection of outliers, our ``LOOPH'' loss method demonstrates improved uncertainty quantification, where smaller variances better represent the true uncertainty in the data. Further, the Hybrid sampling method with the ``LOOPH'' loss function emerges as the most favorable approach for the Ozone dataset with $10\%$ outliers generated into it. This method yields the lowest values across multiple metrics, including RMSE, MDV, and median CI Size. These findings specifically recommend adopting the Hybrid sampling method with the ``LOOPH'' loss function for optimal model performance on this dataset.

\conclusions  
In this study, we investigated the behavior and robustness of GP regression models, particularly focusing on a scalable GP algorithm called \textit{MuyGPs}, when confronted with outlier-affected spatial datasets. We proposed a novel leave-one-out pseudo-Huber (LOOPH) loss method and a down-sampling strategy to enhance the algorithm's robustness and improved prediction capability. Our numerical studies, conducted on both simulated and real-world datasets, provided valuable insights into the capabilities of \textit{MuyGPs} in handling outliers and improving the reliability of GP regression models.

The simulation experiments revealed that \textit{MuyGPs}, when featuring the ``LOOPH'' loss method, maintains low RMSE, small MDV, and accurate confidence intervals even in the presence of extreme observations. Additionally, the down-sampling approach further improved the model's robustness and predictive capabilities, especially when dealing with outlier-affected data, highlighting its potential as a powerful tool for mitigating the adverse effects of unusual observations.

Analyzing real-world U.S. ozone data from LA in 1988, we observed that \textit{MuyGPs}, using the ``LOOPH'' loss method, provides accurate predictions and uncertainty quantification, even when outliers are present. The down-sampling strategy reinforced the algorithm's robustness, making it an attractive choice for applications involving large spatial datasets with potential outliers.

Our study underscores the importance of considering the impact of outliers when employing GP regression models and highlights the potential of the \textit{MuyGPs} algorithm, especially when featuring the proposed ``LOOPH'' loss method and down-sampling techniques. These tools offer practitioners a means to maintain predictive accuracy and reliable uncertainty quantification, even in challenging and large spatial data scenarios. Overall, this work contributes to advancing the understanding of GP regression in the spatial context and offers practical solutions to enhance its applicability in the presence of outliers in the large spatial data regime.
\codeavailability{All codes producing results in this paper can be accessed through \href{https://github.com/Juliettengango1/Robust_GP}{https://github.com/JulietteMukangango}} 

\dataavailability{The maximum daily eight-hour average ozone data can be accessed through the US EPA’s AirData website; \href{https://www.epa.gov/outdoor-air-quality-data/download-daily-data}{www.epa.gov}. The meteorological  data can be accessed directly through \href{https://www.ncei.noaa.gov/access/metadata/landing-page/bin/iso?id=gov.noaa.ncdc:C00516}{www.ncei.noaa.gov}. } 






\appendix
\section{Simulation Results}
Below we illustrate all the metrics computed to evaluate different GP models during the simulation study 
\subsection{ Non-Outlying Results}
\begin{table}[H]
    \centering
    \begin{tabular}{|p{13mm}|p{12mm}|p{5mm}|p{10mm}|p{10mm}|p{10mm}|p{10mm}|p{10mm}|p{10mm}|}
    \hline
       Method  & Loss & $\nu$ & RMSE & CRPS & MAD & MDV & Median CI Size & Cover- age\\
       \hline
        \multirow{6}{5em}{Regular Sampling} & LOOL & $0.1$ & $0.59$ & $0.339$ & $0.415$ &$0.349$ & $2.31$ & $0.94$\\
        && $0.5$ & $0.089$ & $0.051$ & $0.06$& $0.0077$ & $0.344$ & $0.948$\\
        && $1.0$ & $0.009$ &$0.005$ & $0.0056$ & $8.0e^{-5}$ & $0.034$ & $0.945$\\
        \cline{2-9}
        & LOOPH & $0.1$ & $0.581$ & $0.346$ & $0.413$ & $0.342$ & $2.293$ & $0.949$\\
        && $0.5$ & $0.0887$ & $0.0506$ & $0.057$ &$0.007$ & $0.336$ & $0.942$\\
        && $1.0$ & $0.0092$ & $0.0053$& $0.006$ & $6.8e^{-5}$ & $0.032$ & $0.921$\\
        \hline
        \multirow{6}{5em}{Hybrid} & LOOL & $0.1$ & $0.612$ & $0.344$& $0.413$ &$0.344$ & $2.29$ & $0.934$\\
        && $0.5$ & $0.095$ & $0.049$& $0.06$ & $0.0078$ & $0.347$ & $0.934$\\
        && $1.0$ & $0.0098$ & $0.005$ &$0.0061$ &$0.0001$ & $0.035$ & $0.939$\\
        \cline{2-9}
        & LOOPH & $0.1$ & $0.6002$ & $0.344$& $0.4078$ &$0.319$ & $2.216$ & $0.927$\\
        && $0.5$ & $0.089$ & $0.053$ & $0.064$ & $0.0068$ & $0.324$ & $0.931$\\
        && $1.0$ & $0.0096$ & $0.0052$& $0.0063$ &$7e^{-5}$ & $0.0326$ & $0.923$\\
        \hline
        \multirow{6}{5em}{Down-Sampling} & LOOL & $0.1$ & $0.611$ & $0.377$ & $0.402$ &$0.397$ & $2.47$ & $0.959$ \\
        && $0.5$ & $0.089$ & $0.053$& $0.067$ &$0.0082$ & $0.355$ & $0.957$\\
        && $1.0$ & $0.0096$ & $0.0053$ & $6.45e^{-3}$ & $0.0001$ & $0.038$ & $0.955$\\
        \cline{2-9}
        & LOOPH & $0.1$ & $0.603$ & $0.332$ & $0.4023$ & $0.329$ & $2.25$ & $0.942$\\
        && $0.5$ & $0.089$ & $0.053$ & $0.063$ & $0.0075$ & $0.34$ & $0.943$\\
        && $1.0$ & $0.0095$ & $0.0051$ & $0.0059$ & $8e^{-5}$ & $0.035$ & $0.939$ \\
        \hline
        \multirow{3}{5em}{Conventi- onal GP} & NLL & $0.1$ & $0.017$ & $0.054$ & $0.063$ & $0.475$ & $2.703$ & $0.957$\\
        && $0.5$ & $0.003$ & $0.004$ & $0.059$ & $0.883$ & $3.683$ & $0.9502$\\\
        && $1.0$ & $0.0005$ & $0.0027$ & $0.0036$ & $0.523$ & $2.836$ & $0.968$\\
        \hline
        \multirow{3}{5em}{LAGP} & NLL & $0.1$ & $0.947$ & $0.387$ & $0.625$ & $0.704$ & $3.2895$ & $0.978$\\
        && $0.5$ & $0.958$ & $0.381$ & $0.592$ & $0.214$ & $1.812$ & $0.972$\\
        && $1.0$ & $0.725$ & $0.305$ & $0.365$& $0.032$ & $0.703$ & $0.905$\\
        \hline
        \multirow{3}{5em}{Student-t GP} & NLL & $0.1$ & $0.944$ & $0.481$ & $0.626$ & $0.456$ & $5.82$ & $0.975$ \\
        && $0.5$ & $0.747$ & $0.381$ & $0.592$ & $0.385$ & $5.099$ & $0.932$\\
        && $1.0$ & $0.349$ & $0.297$ & $0.404$ & $0.225$ & $3.875$ & $0.948$\\
        \hline
    \end{tabular}
   
   \caption{Results for model evaluation using RMSE, CRPS, MAD, MDV, Median CI size and Coverage. The $\nu$ column represents the true $\nu$ values under which the simulated data were generated.}
    \label{tab:t5}
\end{table}

\subsection{Outlying Results}
\begin{table}[H]
    \centering
    \begin{tabular}{|p{13mm}|p{12mm}|p{5mm}|p{10mm}|p{10mm}|p{10mm}|p{10mm}|p{10mm}|p{10mm}|}
    \hline
       Method  & Loss & $\nu$ & RMSE & CRPS & MAD & MDV &  CI Size & Cover- age\\
       \hline
        \multirow{6}{5em}{Regular Sampling} & LOOL & $0.1$ & $0.625$ & $0.348$ & $0.415$ &$0.569$ & $2.957$ & $0.985$\\
        && $0.5$ & $0.206$ &$0.097$ &$0.087$ & $0.243$ & $1.93$ & $0.998$ \\
         && $1.0$ & $0.21$ &$0.094$ & $0.056$ &$0.278$ & $2.065$ & $0.999$\\
         \cline{2-9}
         & LOOPH & $0.1$ & $0.618$ & $0.345$ & $0.405$ &$0.475$ & $2.704$ & $0.974$\\
         && $0.5$ & $0.126$ & $0.087$& $0.073$ &$0.044$ & $0.819$ & $0.994$\\
         && $1.0$ & $0.058$ & $0.081$ & $0.025$ &$0.031$ & $0.685$ & $0.98$\\
         \hline
         \multirow{6}{5em}{Hybrid} & LOOL & $0.1$ & $0.6012$ & $0.355$& $0.413$ &$0.527$ & $2.846$ & $0.985$\\
         && $0.5$ & $0.176$ & $0.095$& $0.082$ &$0.119$ & $1.35$ & $0.998$\\
         && $1.0$ & $0.227$ & $0.093$& $0.038$ &$0.292$ & $2.118$ & $1.0$\\
         \cline{2-9}
         & LOOPH & $0.1$ & $0.606$ & $0.354$ & $0.422$ &$0.468$ & $2.682$ & $0.971$\\
         && $0.5$ & $0.155$ & $0.079$ & $0.0787$ &$0.061$ & $0.971$ & $0.985$\\
         && $1.0$ & $0.085$ & $0.0743$ & $0.021$ &$0.023$ & $0.593$ & $0.988$\\
         \hline
         \multirow{6}{5em}{Down-Sampling} & LOOL & $0.1$ & $0.619$ & $0.348$ & $0.408$ &$0.488$ & $2.74$ & $0.976$\\
         && $0.5$ & $0.143$ & $0.076$ & $0.0835$ &$0.067$ & $1.016$ & $0.984$\\
         && $1.0$ & $0.0587$ & $0.074$ & $0.037$ &$0.014$ & $0.466$ & $0.988$\\
         \cline{2-9}
         & LOOPH & $0.1$ & $0.6011$ & $0.348$ & $0.412$ &$0.506$ & $2.78$ & $0.985$\\
         && $0.5$ & $0.15$ & $0.086$&$0.078$ & $0.055$ & $0.922$ & $0.987$\\
         && $1.0$ & $0.062$ & $0.075$ & $0.047$ &$0.016$ & $0.503$ & $0.976$\\
         \hline
         \multirow{3}{5em}{Conventi- onal GP} & NLL & $0.1$ & $0.941$ & $0.672$ & $0.904$ & $0.869$ & $3.656$ & $0.992$\\
         && $0.5$ & $0.998$ & $0.556$ & $0.632$ & $1.609$ & $4.972$ & $0.998$\\
         && $1.0$ & $0.869$ & $0.4002$ & $0.443$ & $0.968$ & $3.857$ & $0.9989$\\
         \hline
         \multirow{3}{5em}{LAGP} & NLL & $0.1$ & $1.027$ & $0.534$ & $0.687$ & $2.697$ & $6.438$ & $0.985$\\
         && $0.5$ & $1.14$ & $508$ & $0.587$ & $3.958$ & $7.799$ & $0.997$\\
         && $1.0$ & $0.998$ & $0.463$ & $0.422$ & $3.217$ & $7.03$ & $0.986$\\
         \hline
         \multirow{3}{5em}{Student-t GP} & NLL & $0.1$ & $0.9411$ & $0.6702$ & $0.891$ & $0.846$ & $7.92$ & $0.997$\\
         && $0.5$ & $0.927$ & $0.56$ & $0.651$ & $0.885$ & $8.85$ & $0.988$\\
         && $1.0$ & $0.869$ & $0.384$ & $0.407$ &$0.769$ & $7.465$ & $0.985$\\
         \hline
    \end{tabular}
    \caption{Results for model evaluation using RMSE, CRPS, MAD, MDV, Median CI size and Coverage for data with $10\%$ outliers in it. The $\nu$ column represents the true $\nu$ values under which the simulated data were generated. }
    \label{tab:t6}
\end{table}
\section{U.S Ozone data Analysis Results}
Here we report the results obtained from fitting different \textit{MuyGPs} models for the U.S Ozone data.
\begin{table}[H]
    \centering
    \begin{tabular}{|p{15mm}|p{12mm}|p{10mm}|p{10mm}|p{13mm}|p{13mm}|p{10mm}|p{10mm}|}
    \hline
       Method  & Loss & RMSE & CRPS & MAD & MDV & CI Size & Cover- age\\
       \hline
        \multirow{2}{5em}{Regular Sampling} & LOOL & $0.034$ & $\mathbf{0.019}$ & $0.025$ & $8.6e^{-4}$ & $0.115$ & $0.917$\\
         & LOOPH & $0.0336$ & $0.023$ & $0.0251$ & $\mathbf{7e^{-5}}$ & $\mathbf{0.033}$ & $0.968$\\
         \hline
         \multirow{2}{5em}{Hybrid} & LOOL & $0.0336$ & $0.0209$ & $0.0252$ & $3.51e{-3}$ & $0.232$ & $0.939$\\
         & LOOPH & $0.0336$ & $0.0191$ & $0.025$ & $0.001$ & $0.127$ & $\mathbf{0.956}$\\
         \hline
         \multirow{2}{5em}{Down-Sampling} & LOOL & $\mathbf{0.033}$ & $0.0199$ & $0.0247$ & $0.0005$ & $0.086$ & $0.961$\\
         & LOOPH & $0.0339$ & $0.0232$ & $\mathbf{0.02466}$ & $9e^{-5}$ & $0.0379$ & $0.979$\\
         \hline
    \end{tabular}
    \caption{Results for model evaluation using RMSE, CRPS, MAD, MDV, Median CI size and Coverage.}
    \label{tab:t7}
\end{table}
\begin{table}[H]
    \centering
    \begin{tabular}{|p{15mm}|p{12mm}|p{10mm}|p{10mm}|p{13mm}|p{13mm}|p{10mm}|p{10mm}|}
    \hline
       Method  & Loss & RMSE & CRPS & MAD & MDV & CI Size & Cover- age\\
       \hline
        \multirow{2}{5em}{Regular Sampling} & LOOL & $0.5412$ & $0.384$ & $0.389$ & $1.599$ & $4.95$ & $0.98$\\
         & LOOPH & $0.489$ & $0.283$ & $0.341$ & $0.114$ & $1.323$ & $0.983$\\
         \hline
        \multirow{2}{5em}{Hybrid} & LOOL & $0.54$ & $0.448$ & $0.389$ & $2.611$ & $6.334$ & $0.98$\\
         & LOOPH & $\mathbf{0.489}$ & $0.286$ & $0.342$ & $\mathbf{0.094}$ & $\mathbf{1.204}$ & $0.993$\\
         \hline
         \multirow{2}{5em}{Down-Sampling} & LOOL & $0.556$ & $0.322$ & $0.407$ & $0.149$ & $1.51$ & $0.936$\\
         & LOOPH & $0.509$ & $\mathbf{0.282}$ & $\mathbf{0.268}$ & $0.156$ & $1.548$ & $\mathbf{0.953}$\\
         \hline
    \end{tabular}
    \caption{Results for model evaluation using RMSE, CRPS, MAD, MDV, Median CI size and Coverage for Ozone data with $10\%$ outliers generated in it.}
    \label{tab:t8}
\end{table}








\authorcontribution{Juliette Mukangango conducted all simulations and data analyses and authored the paper. Amanda Muyskens and Benjamin W. Priest contributed to the paper, providing editorial support, guidance throughout the project, and are the creators of the MuyGPs algorithm. Benjamin W. Priest integrated the developed methods into the MuyGPs algorithm.} 

\competinginterests{The authors have no competing interests to declare.} 


\begin{acknowledgements}
This work was performed under the auspices of the U.S. Department of Energy by Lawrence Livermore National Laboratory under Contract DE-AC52-07NA27344 with IM release number LLNL-JRNL-860785 and was supported by the LLNL-LDRD Program under Project No. 22-ERD-028.
We extend our sincere gratitude to Lawrence Livermore National Laboratory for their invaluable support and resources throughout the course of this research. Their commitment to scientific excellence and dedication to advancing knowledge have been instrumental in the success of this study. We are deeply appreciative of their contributions to our work.
\end{acknowledgements}







\bibliographystyle{copernicus}
\bibliography{example.bib}

\end{document}